\documentclass[12pt]{article}

        \usepackage{amsfonts}
        \usepackage{amssymb}

\def\be{\begin{eqnarray}}
\def\ee{\end{eqnarray}}
\def\bee{\begin{eqnarray*}}
\def\eee{\end{eqnarray*}}

\newtheorem{thm}{Theorem}

\newtheorem{defn}[thm]{Definition}

        \def\tr{\rm {Tr}}
        
      \def\Hil{{\mathcal H}}
 \def\cale{{\mathcal E}}
 \def\calm{{\mathcal M}}
\def\cd{{\mathcal D}}

\def\shan{{\rm Shan}}
\def\holv{{\rm Holv}}

\def\wh{\widehat}
\def\wt{\widetilde}

\def\bra{\langle}
\def\ket{\rangle}
\def\kb{ \ket \bra }
\def\dg{\dagger}
\def\ot{\otimes}

\def\dtsig{{\mathbf \cdot \sigma}}

\def\bw{{\bf w}}
\def\nl{\newline}

          \title{Capacity of
Quantum Channels  \\ Using Product Measurements}

        \author{Christopher King \thanks{Partially supported by National
Science
        Foundation Grant DMS-97-05779 and RSDF award
from Northeastern University} \\
Department of
        Mathematics \\ Northeastern University \\
 Boston, MA 02115  \\ {\normalsize king@neu.edu}  \and
Mary Beth Ruskai\thanks{Partially supported by National Science
        Foundation Grant DMS-97-06981 and Army Research Office Grant
   DAAG55-98-1-0374} \\ Department of Mathematics \\
$~~~~~~~~~~~~~~$ University of Massachusetts  Lowell $~~~~~~~~~~~~~~$ \\
Lowell,  MA  01854 USA \\ {\normalsize bruskai@cs.uml.edu}}

 \date{\today \\ ~~ \\
{\it dedicated to Robert Schrader and Ruedi Seiler \\
on the occasion of their 60th birthdays}}

\begin{document}

 \maketitle


\begin{abstract}

The capacity of a quantum channel for transmission of classical
information depends in principle on whether product  states
or entangled states are used at the input, and whether
product or entangled measurements are used at the output.
We show that when {\it product measurements} are used, the capacity of
the channel is achieved with {\it product input states}, so that
entangled inputs do not increase capacity. We show that this result
continues to hold if sequential measurements are allowed, whereby the
choice of successive measurements may depend on the results of
previous measurements.

We also present a new simplified expression which gives an
upper bound for the  Shannon capacity of a channel, and
which  bears a striking
resemblance to the well-known Holevo bound.

\end{abstract}

\tableofcontents



\section{Introduction}

\subsection{Overview}

Bennett and Shor \cite{BS} note that there are, in principle,
four basic types of channel capacities for ``classical'' communication
using quantum signals, i.e., communications in which signals are
sent using an ``alphabet'' of pure states of quantum systems and
decoded using measurements on the (possibly mixed state) signals
which arrive.  The mixed states are the result of noise which
is represented by a {\em stochastic} or completely positive,
trace-preserving map $\Phi$.  The four possible capacities correspond
to using product or entangled states at the input, and using
product or entangled measurements at the output. These are
denoted as follows:
\begin{itemize}
\item[$~~~$] $C_{PP}~~$ product signals and product measurements
\item[$~~~$] $C_{PE}~~$ product signals and entangled measurements
\item[$~~~$] $C_{EP}~~$ entangled signals and product measurements
\item[$~~~$] $C_{EE}~~$ entangled signals and entangled measurements
\end{itemize}
In more precise language ``using product'' means restricting to
products and ``using entangled'' means using arbitrary (product
or entangled) states or measurements.  Hence, it is evident
that $C_{PP} \leq \{ C_{EP}, C_{PE} \} \leq C_{EE}$.  The
main purpose of this note is to show that $C_{PP} = C_{EP}$,
i.e., that if one is restricted to using product measurements,
then using entangled inputs does not increase the capacity.
Thus $C_{PP} = C_{EP} \leq C_{PE} \leq C_{EE}$. It is known
\cite{Fu,Hv1,Hv2} that one can have strict inequality in
$C_{PP} <  C_{PE}$  for certain non-unital channels.  The
question of whether or not one can have strict inequality in
$C_{PE} \leq C_{EE}$ is open, although numerical evidence
\cite{AHW,Smol} suggests equality.

\subsection{Notation and Definitions}

To give precise definitions, we use
 relatively standard notation in which
${\cal M} = \{E_b\}$ denotes a ``positive operator valued measurement''
(POVM) i.e., $E_b > 0$ and $\sum_b E_b = I$.  Let $\rho_j$ denote a
set (or alphabet) of pure state density matrices,  $\pi_j$ a discrete
probability vector, and $\rho = \sum_j \pi_j \rho_j$.
We let ${\cal E} = \{{\pi}_j, {\rho}_j\}$ denote this ensemble of input
states. Both $E_b$ and $\rho_j$ are operators on a Hilbert space $\Hil$,
so that the stochastic map $\Phi$ (representing the noise in the
channel) acts on $B(\Hil)$, the
algebra of bounded operators on $\Hil$. We will write
$\wt{\cale} = \{{\pi}_j, {\Phi(\rho}_j)\}$ for the ensemble of
output states emerging from the channel.

We write the dual of $\Phi$ (or adjoint with respect to the Hilbert-Schmidt
inner product) as $\wh{\Phi}$ so that
$\tr \, [\Phi(\rho) \, E] = \tr \, [\rho \, \wh{\Phi}(E)]$.
The adjoint of a stochastic map takes a POVM
${\cal M} = \{E_b\}$ to another
POVM $\wh{{\cal M}} = \{\wh{E}_b\}$ since the trace-preserving
condition on $\Phi$ is equivalent to $\wh{\Phi}(I) = I$.

\medskip
The information content of a noiseless quantum channel with a fixed input
ensemble and a fixed POVM can be described using the standard Shannon
formula of classical information theory.

\begin{defn} For a fixed ensemble ${\cal E} = \{{\pi}_j,{\rho}_j\}$
and a POVM
${\cal M} = \{E_b\}$ on  a Hilbert space $\Hil$, the quantum
mutual information is given by
\be\label{eq:IqEM}
  I^q( {\cal E}; {\cal M}) =
  S(\tr [\rho E_b]\,) - \sum_j \pi_j  S( \tr [\rho_j E_b]\,) ,
\ee
where $S( \tr [\rho E_b])$ denotes the Shannon entropy
$ - \sum_b p_b \log p_b$ of the
probability vector with elements $p_b = \tr [\rho E_b]$
(and similarly for $S( \tr [\rho_j E_b]\,)$).
\end{defn}
The information content of a noisy channel defined by the stochastic map
$\Phi$ is obtained from (\ref{eq:IqEM}) by replacing
$\cale$ by the output ensemble $\wt{\cale} = \{{\pi}_j, {\Phi(\rho}_j)\}$.
Alternatively, since
$\tr \, [\Phi({\rho}_j) \, E] = \tr \, [\rho_j \, \wh{\Phi}(E)]$,
we could instead choose to regard the ``noise'' as acting
on the POVM, and obtain the capacity from
(\ref{eq:IqEM}) by replacing ${\cal M}$ by $\wh{{\cal M}}$.
Although this viewpoint
is atypical, it can be useful, as we will see in Section \ref{sect:KRproof}.

\begin{defn}  For a stochastic map $\Phi$, an input ensemble
$\cale = \{{\pi}_j,{\rho}_j\}$ and a POVM $\calm = \{E_b\}$,
the quantum information content is given by
\be \label{eq:IqPhiEM}
  I^q_{\Phi}( {\cal E}; {\cal M}) & = & I^q(\wt{\cale};\calm)
 = I^q({\cale};\wh{\calm}) \\
 & = &   S(\tr [\Phi(\rho) E_b]\,)
    - \sum_j \pi_j  S( \tr [\Phi(\rho_j) E_b]\,) .\nonumber
\ee
\end{defn}

We consider memoryless channels in which multiple uses of
the channel are described by the n-fold tensor product
$\Phi \otimes \Phi \ldots \otimes \Phi$ acting on the
tensor product Hilbert space
$\Hil \otimes \Hil \ldots \otimes \Hil$ which we
denote by $\Phi^{\otimes n}$  and $\Hil^{\otimes n}$ respectively.
This allows us to define the `ultimate' information capacity of
the channel as the asymptotic rate achievable when entangled inputs
and measurements are used.

\begin{defn}  The {\em entangled signals/entangled measurements capacity}
of a quantum channel
is defined as
\begin{eqnarray}\label{ent.all.cap}
  C_{EE}(\Phi) = \lim_{n \rightarrow \infty} \,\frac{1}{n} \,
 \sup_{\cale,\calm}\, I^q_{\Phi^{\otimes n}}(\cale;\calm)
\end{eqnarray}
where the supremum is taken over
{\em all} possible (product or entangled) signals and measurements on
$\Hil^{\otimes n}$.
\end{defn}

To define capacity restricted to product measurements, we write
${\calm}^{\otimes n}$ for a  product POVM of the form
$\{E_{b_1}\otimes E_{b_2} \cdots \otimes E_{b_n}\}$.

\begin{defn}  The {\em entangled signals/product measurements capacity} of
a quantum channel
is defined as
\begin{eqnarray}\label{ent.prod.cap}
  C_{EP}(\Phi) = \lim_{n \rightarrow \infty} \,\frac{1}{n} \,
 \sup_{\cale,\calm^{\otimes n}}\,
    I^q_{\Phi^{\otimes n}}(\cale;\calm^{\otimes n}) .
\end{eqnarray}
\end{defn}

Note that the existence of the limits follows from superadditivity of the
classical capacity.

The capacities $C_{PP}$ and $C_{PE}$ can be similarly defined.
We write $\cale^{\otimes n}$ to denote an ensemble of the form
$\{{\pi}_{j_1,\dots,j_n}, {\rho}_{j_1}\otimes\cdots\otimes{\rho}_{j_n}\}$,
where $\{{\rho}_j\}$ is a fixed collection of states, and
$\{{\pi}_{j_1,\dots,j_n}\}$ is some joint probability distribution.

\begin{defn}  The {\em product signals/entangled measurements capacity} of
a quantum channel
is defined as
\begin{eqnarray}\label{prod.ent.cap}
  C_{PE}(\Phi) = \lim_{n \rightarrow \infty} \,\frac{1}{n} \,
 \sup_{\cale^{\otimes n},\calm}\,
    I^q_{\Phi^{\otimes n}}(\cale^{\otimes n};\calm) .
\end{eqnarray}
\end{defn}
\begin{defn}  The {\em product signals/product measurements capacity} of
a quantum channel
is defined as
\begin{eqnarray}\label{prod.prod.cap}
  C_{PP}(\Phi) = \lim_{n \rightarrow \infty} \,\frac{1}{n} \,
 \sup_{\cale^{\otimes n},\calm^{\otimes n}}\,
    I^q_{\Phi^{\otimes n}}(\cale^{\otimes n};\calm^{\otimes n}) .
\end{eqnarray}
\end{defn}

The additivity of classical information capacity immediately
implies the following result.
\begin{thm}  The {\em product signals/product measurements
capacity} of a quantum channel is given by
\begin{eqnarray}\label{Shan.cap}
  C_{PP}(\Phi) = C_{\shan}(\Phi) = \sup_{{\mathcal E}, {\cal M}} \,
I^q_{\Phi}( {\cal E}; {\cal M}) .
\end{eqnarray}
which we call the {\em Shannon  capacity}.
\end{thm}

A far deeper result is that $C_{PE}(\Phi)$ can be re-expressed in terms
of the well-known Holevo bound \cite{Hv0,Hv1,NC}. This result was
proved independently in \cite{Hv1} and \cite{SW}, building on
earlier work in \cite{Hv2} and \cite{HJSWW}.
\begin{thm} {\em (Holevo-Schumacher-Westmoreland)}
The {\em product signals/entangled measurements
capacity}  of a quantum channel is given by
\begin{eqnarray}\label{Holv.cap}
  C_{PE}(\Phi) = C_{\holv}(\Phi) = \sup_{{\mathcal E}} \left(
  S[ \Phi(\rho)] - \sum_j \pi_j  S[ \Phi(\rho_j) ] \right)
\end{eqnarray}
where $S(P) = -\tr \, (P \,\log P)$ denotes the von Neumann entropy
of the density matrix $P$. We call this the {\em Holevo capacity}
of the channel.
\end{thm}

\subsection{Summary of Results}

Our main result, that using entangled inputs with product measurements
does not increase the capacity of a channel, can be stated as
\begin{thm}\label{thm.EP}
For any stochastic map, $C_{EP}(\Phi) = C_{\shan}(\Phi)$.
\end{thm}

There is another implementation of product measurements which has
the potential for
a greater capacity. It involves a sequence of POVM's on the product spaces
${\Hil}^{\otimes n}$, whereby
the POVM for the second measurement depends on the result of the
first measurement, the POVM for the third measurement depends on the
results of the first
two measurements, and so on. The idea is that ``Bob'' can choose his
successive POVM's
based on the results of previous measurements. We  write $C_{EP}^{\rm
cond}(\Phi)$ for the maximum asymptotic rate achievable for such a
sequence of conditional POVM's, with entangled inputs allowed. (The
precise definition of a conditional POVM is postponed to
Section \ref{sect:KRproof} and the capacity is given by
(\ref{def:Iqcond}).)   Our next result
shows that using such conditional POVM's with entangled inputs again does
not increase the channel capacity.
\begin{thm}\label{thm:cond}
For any stochastic map, $C_{EP}^{\rm cond}(\Phi) = C_{\shan}(\Phi)$.
\end{thm}
Theorem \ref{thm:cond} was proved independently (and simultaneously),
using different methods, by P. Shor \cite{Shor}, and also later proved
independently by A. Holevo \cite{Hv4}.
 A conditional POVM is not the most general situation
involving product measurements, which would be a POVM in which each
measurement can be written as a tensor product.  Except for the obvious
bounds, we know of no results for the  capacity associated with such
POVM's.
\medskip

The capacity of a classical channel  can be written as the
(suitably restricted) supremum of the classical mutual information.
We extend this observation  to the quantum case, using a
tensor product formulation whereby
the first two (and possibly all four) of these basic
capacities are realized using mutual information in the
form of the relative entropy of a density matrix and the product of its
reduced density matrices. This leads to the following upper bound.
\begin{thm} \label{thm:upbd}For any stochastic map,
\bee
  C_{EP}(\Phi) \leq \sup_{\calm, \rho}  \left[ S(\rho)  -
\sum_b S\left(\sqrt{\rho}\, \wh{\Phi}(E_b) \sqrt{\rho}\right) +
S(\tau)  \right]
\eee
where $\tau_b =  \tr \,[ \Phi(\rho) \, E_b]  = 
\tr \,[\rho \, \wh{\Phi}(E_b)]$.
\end{thm}
We call the quantity on the right $U_{EP}$, and we conjecture that
it is equal to $C_{EP}$, i.e. that equality holds in Theorem
\ref{thm:upbd}  above.  We motivate and study $U_{EP}$ in Section
\ref{sect:UEP} where we show that it can  be rewritten in a form similar to
the Holevo capacity. Combined with Theorem \ref{thm.EP} above,
 this conjectured equality would provide a simplified expression
for the Shannon capacity of any channel,
whereby the sup over both input ensemble and POVM is replaced by a sup
over {\it one} average input state and the POVM.


Although the proof of Theorem \ref{thm:cond} does not depend
on our tensor product reformulation, we present this material first, in
the following section, because we feel it gives some useful
insights.  Section \ref{sect:mut.info} is largely pedagogical
and provides the motivation for our conjectured expression for $C_{EP}$.
Section \ref{sect:QC.chan} is also primarily pedagogical; it
introduces the reader to Holevo's C-Q and Q-C channels \cite{Hv1}.
This leads to a short proof of both the well-known
Holevo bound and the new bound in Theorem \ref{thm:upbd}.
Moreover, the additivity of Q-C channels implies Theorem \ref{thm.EP}
and motivates our proof of Theorem \ref{thm:cond}. The reader primarily
interested in this proof can skip directly to Section \ref{sect:KRproof}.



\section{Capacity from Mutual Information} \label{sect:mut.info}

\subsection{Classical background}

The {\em classical mutual information} of two
random variables $X$ and $Y$
measures how much information they have in common and is
given by
\be \label{eq:mut.info.c}
  I^c(X;Y)   \equiv \sum_{x,y} p(x,y) \log \frac{p(x,y)}{ p(x) p(y)}
\ee
If $X$ and $Y$ represent the input and output distributions
of a channel, then the classical Shannon capacity is the
supremum of $I^c(X;Y)$ taken over all possible joint distributions
allowed by the channel.

The Shannon capacity of a quantum channel can also be obtained
in this way provided that the joint distribution arises from a
quantum communication process $(\Phi, \cale, \calm) $ as
\be \label{eq:dual}
p(j,b) = \pi_j \tr  \, [\Phi(\rho_j) E_b] =
  \pi_j \tr \, [\rho_j \wh{\Phi}(E_b)]
\ee
Although the stochastic map $\Phi$ is usually regarded as noise acting
on the signals $\rho_j$, it is important to recognize that it has
another interpretation corresponding to the second
expression for $p(j,b)$ in (\ref{eq:dual}) above.  In the second
case, the channel transmits  signals faithfully, but the
``noise'' distorts the measurement process by converting
the POVM $\{E_b\}$ to a modified POVM $\{\wh{E}_b = \wh{\Phi}(E_b) \}$
implemented by the action of the dual of $\Phi$.

In order to
make the transition from classical to quantum communication,
it is sometimes useful to consider a classical probability
vector $p(x)$ as the diagonal of a matrix $P$.  We can then
write the {\em relative entropy}
\be \label{eq:relentdef}
  H(P,Q) = \tr [ P \log P - P \log Q ]
\ee
in a form which reduces to the usual classical expression when
$P$ and $Q$ are diagonal, but is also valid when $P$ and $Q$
are density matrices representing mixed quantum states.
In this notation (\ref{eq:mut.info.c}) becomes
\be \label{eq:mutinf.relent}
  I^c(X;Y) = H[P_{12}, P_1 \otimes P_2]
\ee
where $P_{12}, P_1$, and $P_2$ are diagonal matrices with
non-zero entries $p(x,y), p(x)$ and $p(y)$ respectively.


\subsection{Tensor Product Reformulation}

A reformulation and generalization of mutual information
and capacity can be made using {\em formal} tensor products.
It should be emphasized that
this is done for convenience of notation and is {\em distinct}
from the tensor products used in describing multiple uses of
the channel. Let
$\Hil_{ABQR} = {\bf C}^J \otimes {\bf C}^M \otimes \Hil \otimes \Hil $
where $j = 1 \ldots J$,  $b = 1 \ldots M$ and
$\Hil_Q = \Hil_R = \Hil$ is the original Hilbert space on which $\rho$
and $E_b$ act. The partial traces then correspond to $T_A = \sum_j$,
$T_B = \sum_b$, $T_Q = \tr$, and $T_R = \tr$.

 \bigskip

\noindent Let $P_{ABQ}$ be the block diagonal matrix with blocks
$\pi_j \sqrt{\Phi(\rho_j)} E_b \sqrt{\Phi(\rho_j)}$ and
\nl $~~~\wh{P}_{ABQ}$ the block diagonal matrix with blocks
$\pi_j \sqrt{\rho_j} \, \wh{\Phi}\, (E_b) \sqrt{\rho_j}$.
\medskip

\noindent Then  $P_{AB} \equiv T_Q P_{ABQ} = T_Q \wh{P}_{ABQ} \equiv
\wh{P}_{AB}$ and
\begin{itemize}
\item[$P_{AB}$]  is a diagonal matrix with
(non-zero) elements  $p(j,b) = \pi_j  \tr \, [\Phi(\rho_j) \, E_b]$,
\item[$P_A~$]   $\equiv T_{BC} P_{ABQ} = T_B P_{AB}$
is a diagonal matrix with  elements $\delta_{ij} \pi_j$,
\item[$P_B~$]   $\equiv T_{AQ} P_{ABQ} = T_A P_{AB}$
is a diagonal matrix with elements  $\delta_{ab} \tau_b$
where \linebreak $\tau_b =  \tr  \Phi(\rho) E_b  =  \tr \rho
\wh{\Phi}(E_b)$ as in Theorem \ref{thm:upbd}.
\end{itemize}

It is straightforward to verify that
\be \label{mut.shan}
  C_{PP} \equiv C_{\shan}(\Phi) & = & \sup_{\cale,\calm} \left[
S(P_B) - S(P_{AB}) + S(P_A) \right] \label{shan.tens}   \\
 & = & \sup_{\cale,\calm} H(P_{AB}, P_A \otimes P_B) =
   \sup_{\cale,\calm} I^{q}_{\Phi}(\cale;\calm)  \nonumber \\
 & = &  \sup_{\cale,\calm} I^q(\wt{\cale}; \calm)   =
\sup_{\cale,\calm} I^q(\cale; \wh{\calm}).   \label{mut.shan.supp}
\ee
where the last line in (\ref{mut.shan.supp}), although redundant
is included to emphasize the fact that we can suppress the
explicit dependence on $\Phi$ by using either a restricted ensemble
with $\wt{\rho}_j = \Phi(\rho_j)$ or a restricted
POVM of the form $\wh{\Phi}(E_b)$.

Note that all the matrices in (\ref{mut.shan}) above are diagonal
and could be replaced by probability vectors.  The quantum
character of the channel is hidden in the fact that $P_{AB}$
must be the reduced density matrix of a $P_{ABQ}$ of the
form above with quantum blocks.  Thus we might have replaced
$\sup_{\cale,\calm}$ above by either
$\sup_{P_{ABQ}} H(P_{AB}, P_A \otimes P_B) $ or
$\sup_{\wh{P}_{ABQ}} H(P_{AB}, P_A \otimes P_B) $ with the
understanding that the supremum was to be taken over those
$P_{ABQ}$ or $\wh{P}_{ABQ}$ with the block diagonal form
given above.

 \bigskip


We can find a similar expression for the Holevo capacity  by noting that
\begin{itemize}
\item[$P_{AQ}$]   $\equiv T_B P_{ABQ}$ is a block diagonal matrix with
blocks $\pi_j \Phi(\rho_j)$, and
\item[$P_Q~$]   $\equiv T_{AB} P_{ABQ} = T_A P_{AQ} = \Phi(\rho)$.
\end{itemize}
It is again straightforward to verify that
\be
   C_{PE} \equiv C_{\holv}(\Phi) & = & ~ \sup_{\cale} ~\left[
S(P_Q) - S(P_{AQ}) + S(P_A) \right] \label{holv.tens}   \\
& = &  ~ \sup_{\cale}~ H(P_{AQ}, P_A \otimes P_Q) . \nonumber
\ee
We can interpret this as a classical to quantum
mutual information between the classical probability
distribution $\pi_j$ of the input alphabet and the average
quantum distribution  $\Phi(\rho)$ which emerges from the channel.

\medskip

We conclude by observing that the entanglement assisted capacity
of \cite{BSST} can be written in a similar way as
\be \label{eq:entang.cap}
   \sup  \left\{ H(\rho_{QR}, \rho_Q \otimes \rho_R) : \rho_{QR} =
   (\Phi \otimes I )( |\Psi \kb \Psi |) \right\}
\ee
with $\Psi \in {\bf C}^2 \otimes {\bf C}^2 $.  This differs slightly
from eq. (4) of \cite {BSST}.  However, because $|\Psi \kb \Psi |$ is pure,
their $S(\rho) = S[T_2 (|\Psi \kb \Psi |)] = S[T_1 (|\Psi \kb \Psi |)] =
 S(\rho_R)$ in our notation.  Thus the expression in (\ref{eq:entang.cap})
above is equivalent to eq. (4) of \cite {BSST}.  This is a form
of quantum to quantum mutual information between the subsystems
of an entangled pair, one of which is subjected to noise via
transmission through the channel.


We also  expect that the capacity $C_{EE}$ can be expressed as
a (different) quantum to quantum mutual information.  Unfortunately
the precise form has eluded us.  This approach does, however,
lead in a natural way to a new expression related to $C_{EP}$.

\subsection{Proposed expression for $C_{EP}$} \label{sect:UEP}

To motivate our new candidate for $C_{EP}$, we let $P_{BR}$ be
the block diagonal matrix with blocks
$\sqrt{\rho}\, \wh{\Phi}(E_b) \sqrt{\rho}$.   Then
\begin{itemize}
\item[$P_B$]  $ \equiv T_R P_{BR}$ is a diagonal matrix with
 elements  $\tau_b$.
\item[$P_R$] $ \equiv T_{B} P_{BR} = \rho $
\end{itemize}
and define
\be \label{eq:UPE}
U_{EP}(\Phi) & = & ~ \sup_{\calm, \rho} ~\left[
S(P_R) + S(P_B) - S(P_{BR}) \right] \label{qc.tens}  \nonumber \\
& = &  ~\sup_{\calm, \rho}~ H(P_{BR}, P_R \otimes P_B) \\
& = & ~\sup_{\calm, \rho}  \left[ S(\rho)  -
\sum_b S\left(\sqrt{\rho}\, \wh{\Phi}(E_b) \sqrt{\rho}\right) +
S(\tau)  \right] \nonumber \\
& = & ~\sup_{\tau_b, \gamma_b}  \left[ S(\gamma)  - \sum_b \tau_b
S(\gamma_b)\right]
 \label{eq:CPE.alt}
\ee
where
$\gamma_b = \frac{1}{\tau_b} \sqrt{\rho}\, \wh{\Phi}(E_b) \sqrt{\rho}$
and $\gamma = \sum_b \tau_b \gamma_b = \rho$.
The last form (\ref{eq:CPE.alt}),  looks like the
Holevo capacity with the input ensemble $\cale = \{ \pi_j, \rho_j \}$
replaced by a new ``output measurement ensemble''
$\{ \tau_b, \gamma_b \}$.
 How can we characterize this ensemble?
Using Kraus operators we can write
$\Phi(\rho) = \sum_k A_k^{\dg} \rho A_k$, where
$\sum_k  A_k A_k^{\dg} = I$. It follows that
$\gamma_b = \sum_k B_k^{\dg} E_b B_k$ with  $B_k = A_k^{\dg} \sqrt{\rho}$.
Hence $\gamma_b$ is a density matrix in the range of a completely
positive map which, rather than being trace-preserving or unital,
satisfies $\sum_k  B_k B_k^{\dg} = \Phi(\rho)$.  If we
define
$\Gamma_\rho(P) = \sqrt{\rho} \, \wh{\Phi}(P) \sqrt{\rho} $ we
can write
\be
U_{EP}(\Phi) =   \sup_{ \rho, \calm} \left(
  S \big[ \Gamma_\rho(I) \big] - \sum_b \tau_b 
       S \big[ \Gamma_\rho \big(\tau_b^{-1} E_b \big) \big] \right).
\ee
A different characterization is given
in the next section as a condition on $P_{BR}$.

We can interpret (\ref{eq:UPE}) as a  quantum  to classical
mutual information between the average input $\rho$ and
the classical probability vector $\tau_b$ associated with
the correspondingly averaged output measurements
$ \tr\,  [\rho \, \wh{\Phi}(E_b)]$.

We conjecture that $U_{EP} = C_{EP}$ although we
can only show $U_{EP} \geq C_{EP}$, which is proved in the
next section.
Note  if $\Phi$ is the completely noisy
channel which maps every density matrix  to the identity, then
 $P_{BR} = P_B \otimes P_R$ so that
$H(P_{BR}, P_B \otimes P_R) = 0$ as expected.  This also holds
 if  $\rho$ is a one-dimensional projection.


\subsection{Optimization constraints}

We can rewrite all of these expressions for capacity as the
suitably constrained supremum of an ``Input-Output''
mutual information,
$H(\rho_{\cal IO}, \rho_{\cal I} \otimes \rho_{\cal O})$, i.e.,
\be
  \sup \left\{ H(\rho_{\cal IO}, \rho_{\cal I} \otimes \rho_{\cal O}) :
\rho_{\cal IO} ~\hbox{is a density matrix in}~ X_{\cal IO} \right\}
\ee
where the subset $X_{\cal IO}$ lies in
${\cal A}_{\cal I} \otimes {\cal A}_{\cal O}$
and the algebra ${\cal A}$ is either
${\bf C}^{n \times n}$ or ${\bf D}^n$,
the algebra of diagonal $n \times n$ matrices.  We will let
${\mathcal G} = \{ E : 0 \leq E \leq I \}$ denote the set of
positive semi-definite operators less than the identity,
${\mathcal D}$ the set of density matrices, and  $\leq \cd$
   the set  of positive semi-definite matrices
  with trace $ \leq 1$, i.e., the set of matrices $\lambda P$ where
$P$ is a density matrix and $0 \leq \lambda \leq 1$.
\begin{itemize}

\item[$~$]  $C_{PP}:~ X_{\cal IO} =
 \left\{ \rho_{AB} = \tr_Q \, \rho_{ABQ} :
\rho_{AQ}^{-1/2}\,  \rho_{ABQ} \, \rho_{AQ}^{-1/2}
 \in {\bf D}^n \otimes {\bf D}^n \otimes  \wh{\Phi}({\mathcal G})  \right\}$.
\nl $~~~$ In the case of maps on ${\bf C}^{2 \times 2}$ we  expect this
to be a subset of
\nl $~~~$ ${\bf D}^2 \otimes {\bf D}^2 $ although,
 in principle, it could be a subset of ${\bf D}^4 \otimes {\bf D}^4. $

\item[$~$]  $C_{PE}:~ X_{\cal IO} =\Big\{ \rho_{AQ} :
  \rho_{AQ} \in   {\bf D}^n \otimes  \Phi( \leq \cd)   \Big\}$.

\item[$~$]  $U_{EP}:~ X_{\cal IO} =\left\{ \rho_{BR} :
     \rho_{B}^{-1/2} \, \rho_{BR} \, \rho_{B}^{-1/2} \in
       {\bf D}^n \otimes  \wh{\Phi}({\mathcal G})
  \right\}$

\item[$~$]  $C_{EE}:~ $ We know only that
     $X_{\cal IO} \subset {\bf C}^{n \times n} \otimes {\bf C}^{n \times n}. $
\end{itemize}

In order to conclude that these expressions are equivalent to
those given previously, we need to verify that when
 $\rho_{\cal IO}$ is in the indicated set, one can always find a
corresponding ensemble $\cale$ and/or POVM $\calm$.  The block
diagonal conditions implicit in the notation above and
the fact that
$\Phi$ and $\wh{\Phi}$ are trace-preserving and identity
preserving respectively, makes this quite straightforward.

When $n = 2$, we can describe ${\mathcal G}$ explictly by
writing $E = w_0 I + \bw \dtsig$
where $\sigma = (\sigma_x, \sigma_y, \sigma_z )$ denotes
the formal vector of Pauli matrices and $\bw$ in ${\bf R}^3$.
Then $0 \leq E \leq I$
if and only if $|\bw| \leq \min \{ w_0, 1-w_0 \}$ so that
\bee
 {\mathcal G} = \bigcup_{w_0 \in [0,1]}
\Big\{ E = w_0 I + \bw \dtsig : |\bw| \leq \min \{ w_0, 1-w_0 \} \Big\}.
\eee

\bigskip

\section{Bounds via Q-C Channels} \label{sect:QC.chan}

Holevo \cite{Hv3} introduced an extremely useful family of
stochastic maps of the form
\begin{eqnarray}\label{chan.omega}
\Omega(P) = \sum_k R_k~ \tr(P X_k)
\end{eqnarray}
where $R_k$ is a family of density matrices, $X_k$ is a POVM.
He also distinguished two important subclasses of these
channels
\begin{itemize}
\item[$\Omega_{QC}$] Quantum-classical channels in which
 $R_k = |e_k\kb e_k|$  so that each density matrix
is a one-dimensional projection from an orthonormal basis $\{e_k\}$.
\item[$\Omega_{CQ}$] Classical-quantum channels in which
 $X_k = |e_k\kb e_k|$  so that the POVM is a  partition of unity arising
from an orthonormal basis $\{e_k\}$.
\end{itemize}
Holevo \cite{Hv1} showed that the quantum capacity of such
channels is additive, i.e.,
\bee
C_{PE}(\Phi_{QC} \otimes \Phi_{QC} \ldots \otimes \Phi_{QC}) =
C_{PE}(\Phi_{QC}^ {\, \otimes n}) = n \, C_{PE}(\Phi_{QC})
\eee
and similarly for $C_{PE}(\Phi_{CQ}^ {\, \otimes n}) = n \,
C_{PE}(\Phi_{CQ})$.  In the next section, we use Holevo's strategy for
proving  additivity for $\Phi_{CQ}$ to prove Theorem \ref{thm:cond}.

We now show that both the celebrated ``Holevo bound''
$C_{PP}(\Phi)  \leq   C_{PE}(\Phi)$ and the new bound $C_{PP}(\Phi)
\leq   U_{EP}(\Phi)$ follow easily from the
monotonicity of relative entropy under  $\Omega_{QC}$ channels.
Our strategy is similar to one used earlier by Yuen and Ozawa \cite{YO}.

In the first case, we let $\Omega_{QB}$ be a Q-C map of the form
(\ref{chan.omega}) with $X_b = E_b$ and $R_b =  |e_b\kb e_b|$.  Then
\begin{eqnarray}
 H(P_{AB}, P_A \otimes P_B) & = & \nonumber
H\left[ \Omega_{QB}(P_{AQ}), \Omega_{QB}( P_A \otimes P_Q) \right] \\
 & \leq  & H(P_{AQ}, P_A \otimes P_Q)  \label{eq:Holv.bd}
\end{eqnarray}
where $P_{AQ}$ and $P_{AB}$ are as in Section \ref{sect:mut.info}
and we have suppressed the identity in
$I \otimes \Omega_{QB}$. Taking the supremum over ${\cal E}$ yields
$C_{PP}(\Phi)  \leq   C_{PE}(\Phi)$.

For the new bound,
let $\Omega_{RA}$ be a Q-C map of the form (\ref{chan.omega})
with  $X_j =\pi_j \rho^{-1/2} \rho_j \, \rho^{-1/2}$ and
$R_j =|e_j \kb e_j| $,
so that $ \Omega_{RA}(P_{BR}) = P_{AB}$.  Then
\begin{eqnarray}
  H(P_{AB}, P_A \otimes P_B)  & = &
   H[\Omega_{RA}(P_{BR}),\Omega_{RA}(P_B \otimes P_R)]    \nonumber \\
   & \leq  & H(P_{BR}, P_B \otimes P_R)   \nonumber
\end{eqnarray}
from which it follows that $C_{PP}(\Phi)  \leq   U_{EP}(\Phi)$.

\medskip

\noindent{\bf Remark:}  It may appear that the argument in
(\ref{eq:Holv.bd}) above yields a simple proof of the Holevo
bound without using the strong subadditivity (SSA) of
relative entropy \cite{SSA}
as in  \cite{SWW}.  However, Lindblad \cite{Lind} made the useful
observation that any stochastic map can be represented
as the partial trace after interaction with an auxiliary system, i.e.,
$\Phi(P) = T_B \big[ U_{AB} P \otimes E_B U_{AB}^{\dg} \big] $
In fact, he used this representation to obtain monotonicity
as a corollary of SSA.   Thus, the arguments used to obtain the Holevo
bound via monotonicity (as above or in \cite{YO})  and via SSA (as in
\cite{SWW}) are essentially equivalent.  In the latter approach, an
auxiliary system is added explicitly and then discarded; in the
former, this is done implicitly via Lindblad's representation
theorem.  Further discussion of the history of the closely
connected properties of SSA, monotonicity of relative entropy and the
joint convexity of relative entropy is given in \cite{OP,Rusk,W} .



\section{Proof of Additivity Using
Q-C Channels}\label{sect:KRproof}

Theorem \ref{thm.EP} can be obtained from Holevo's result
\cite{Hv3} that $C_{\holv}(\Omega_{QC})$ is additive, i.e.,
if $\Gamma$ is a Q-C channel of the form
following (\ref{chan.omega}), then $C_{\holv}(\Gamma)$ is additive.
To show how this follows, we define
\be \label{eq:meas.chan}
  \Gamma_{\Phi,\calm}(P) =
       \sum_b |e_b \kb e_b| \,   \tr  \, [P \wh{\Phi}(E_b)] .
\ee
Then $\Gamma_{\Phi,\calm}(P) $ is a Q-C channel with
$X_n = \wh{\Phi}(E_b)$.  Moreover,
$ \sup_{\cale} \,  I^q_{\Phi}( {\cal E}; {\cal M}) =
      C_{\holv} \big(\Gamma_{\Phi,\calm} \big)$, and the additivity
of $C_{\holv} \big(\Gamma_{\Phi,\calm} \big)$ implies
$ \sup_{\cale} \,  I^q_{\Phi^{\ot n}}( {\cal E}; {\cal M}^{\ot n}) =
      C_{\holv} \big(\Gamma_{\Phi,\calm}^{\ot n} \big) = n C_{\holv}
\big(\Gamma_{\Phi,\calm} \big)$. Then Theorem \ref{thm.EP} follows from
\bee
  C_{\holv}(\Phi) = \sup_{\cale,\calm} \,  I^q_{\Phi}( {\cal E}; {\cal M})
= \sup_{\calm} C_{\holv} \big(\Gamma_{\Phi,\calm} \big) .
\eee


In order to prove Theorem
\ref{thm:cond}, we will need to extend Holevo's result.
Our extension, which we present  below, follows Holevo's strategy
\cite{Hv3} with the identity (\ref{eq:mutinfo.ident}) replacing
subadditivity.  This also provides a self-contained proof of
Theorem \ref{thm.EP}, since a product measurement is a special
case of a conditional measurement.

First consider a product
channel with Hilbert space ${\Hil}_1 \otimes {\Hil}_2$ and noise operator
${\Phi}_1 \otimes {\Phi}_2$.
Let $\cale_{12} = \{{\pi}_j, {\rho}_j\}$ be an
ensemble of possibly entangled input states
on ${\Hil}_1 \otimes {\Hil}_2$. Let ${\cal M}_1 = \{E_{b}\}$
denote the POVM on ${\Hil}_1$ which implements the
first measurement, and for each $b$ let
${\cal M}_{2}(b) = \{E_{c}^{(b)}\}$ denote the POVM on ${\Hil}_2$ which
implements the second measurement.  We then define a joint POVM
${\cal M}_{12}$ on ${\Hil}_1 \otimes {\Hil}_2$, namely
$\{E_{b} \otimes E_{c}^{(b)}\}$.  Note that although each element
of $\calm_{12}$ is a product, the joint measurement need not be
the product of independent measurements $\calm_1 \otimes \calm_2$.
This is the result of the fact that the
second measurement may be conditioned on the results of
the first.  Nevertheless, it is easy to verify that $\calm_{12}$
is a POVM since
\bee
\sum_{b,c} E_{b} \otimes E_{c}^{(b)} = \sum_b E_b \otimes
   \Big( \sum_c  E_{c}^{(b)} \Big)
 = \sum_b E_b \otimes I .
\eee

The information content of a channel using such conditioned
measurements is
\be \label{eq:product info}
I^{q}_{\Phi_1 \otimes \Phi_2}(\cale_{12};\calm_{12}) =
I^q(\cale_{12};\wh{\calm}_{12}) = I^q(\wt{\cale}_{12};\calm_{12})
\ee
where $\wh{\calm}_1, \wh{\calm}_2(b)$ and $\wh{\calm}_{12}$ denote
the POVM's in which
$E_b$ is replaced by $F_b = \wh{\Phi_1}(E_b)$ and $E_{c}^{(b)}$ is
replaced by
$F_{c}^{(b)} = \wh{\Phi_2}(E_{c}^{(b)})$, and we have used the
notation defined in (\ref{eq:IqEM}) and  (\ref{eq:IqPhiEM}).
Because we are interested in studying the capacity for a fixed set of
POVM's, we use the form $I^q(\cale_{12};\wh{\calm}_{12})$ and
proceed as if we were considering a noiseless channel with
a restricted POVM of the above form.  Although this viewpoint
is useful, it is not essential.   The argument would work
equally well if we explicitly included the stochastic maps
or used the form $I^q(\wt{\cale}_{12};\calm_{12})$
and defined  reduced density matrices using partial traces
acting on, e.g., $(\Phi_1 \otimes \Phi_2)(\rho_j)$.


For any input ensemble $\cale_{12}$ we now define a
pair of associated input ensembles on ${\Hil}_1$ and
${\Hil}_2$ respectively.  For this purpose it is useful
to let $T_j$ denote the partial trace over ${\Hil}_j$.
First, let ${\rho}_{j}^{(1)} = T_2 \, [{\rho}_{j}]$ be the indicated
reduced density matrix and
${\cal E}_{1} = \{ {\pi}_{j}, {\rho}_{j}^{(1)} \}$.
This is our ensemble on ${\Hil}_1$. Second,
for each $j$ and $b$, define a state on ${\Hil}_2$ by
\be
{\rho}_{j,b}^{(2)} = p(b|j)^{-1} ~
  T_1 \,  [ ({\rho}_{j}) \, (F_{b} \otimes I)],
\ee
where $p(b|j) = \tr \, [ {\rho}_{j} (F_{b} \otimes I)]$. Then the
corresponding input ensemble on ${\Hil}_2$ is \linebreak
${\cal E}_{2}(b) = \{ p(j|b), \,\,{\rho}_{j,b}^{(2)}\}$, where
$p(j|b) = p(b|j) {\pi}_j / p(b)$ and
\be
p(b) = \sum_j {\pi}_j  p(b|j) =
\tr \, \left[ \Big(\sum_j {\pi}_j {\rho}_j \Big)
(F_{b} \otimes I) \right] .
\ee
We claim that
\be \label{eq:mutinfo.ident}
I^q(\cale_{12};\wh{\calm}_{12}) = I^q(\cale_1;\wh{\calm}_1) +
  \sum_b p(b) \, I^q \big[{\cal E}_{2}(b);\wh{\calm}_{2}(b)\big] .
\ee
Since
\be
I^q\big[{\cal E}_{2}(b);\wh{\calm}_{2}(b)\big] =
  I^{q}_{\Phi_2}\big[{\cal E}_{2}(b);{\calm}_{2}(b)\big] \leq
      C_{\shan}(\Phi_2)
\ee
it follows immediately from (\ref{eq:mutinfo.ident}) that
\be\label{ineq:one}
 I^q(\cale_{12};\wh{\calm}_{12}) & \leq & I^q(\cale_1;\wh{\calm}_1)
  + \sum_b p(b) \, C_{\shan}(\Phi_2) \nonumber \\
  & = & I^q(\cale_1;\wh{\calm}_1) +  C_{\shan}(\Phi_2) .
\ee
Taking the supremum over channels of this type, which we now
emphasize by writing $\calm_{12}^{\rm cond}$, gives
\be \label{ineq:two}
\sup_{\cale_{12},\calm_{12}^{\rm cond}}
 \, I^{q}_{\Phi_1 \otimes \Phi_2}(\cale_{12};\calm_{12}^{\rm cond}) & = &
\nonumber
 \sup_{\cale_{12},\wh{\calm}_{12}^{\rm
cond}}I^q(\cale_{12};\wh{\calm}_{12}^{\rm cond}) \\
& \leq & C_{\shan}(\Phi_1) +  C_{\shan}(\Phi_2) .
\ee
However by restricting to product ensembles and product POVM's in the sup
on the left
side of (\ref{ineq:two}), and using additivity of the classical capacity
(\ref{Shan.cap}),
we deduce
\be \label{ineq:three}
\sup_{\cale_{12},\calm_{12}^{\rm cond}}
 \, I^{q}_{\Phi_1 \otimes \Phi_2}(\cale_{12};\calm_{12}^{\rm cond})
 \geq & C_{\shan}(\Phi_1) +  C_{\shan}(\Phi_2) .
\ee
Hence we have equality in (\ref{ineq:two}).

\medskip
Now consider the $n$-fold product channel $\Phi_1 \otimes \cdots \otimes
\Phi_n$. Let ${\cal M}^{\rm cond}$
be a conditional POVM  on ${\Hil}_1 \otimes \cdots \otimes {\Hil}_n$.
By assumption, every operator in this POVM has the form
$E_{b} \otimes E^{(b)}_c$ where $\{E_{b}\}$ is a conditional
POVM ${\cal N}^{\rm cond}$ on
${\Hil}_1 \otimes \cdots \otimes {\Hil}_{n-1}$, and for each $b$,
$E^{(b)}_c$ constitute a POVM on ${\Hil}_n$. Also,
for any input ensemble
$\cale$ on
${\Hil}_1 \otimes \cdots \otimes {\Hil}_n$, let
$\cale'$ be the ensemble of reduced density matrices
on  ${\Hil}_1 \otimes \cdots \otimes {\Hil}_{n-1}$. Then (\ref{ineq:two})
implies
\be \label{ineq:four}
\sup_{\cale,{\cal M}^{\rm cond}}
\lefteqn{ \, I^{q}_{\Phi_1 \otimes \Phi_2 \otimes \cdots \otimes \Phi_n}
(\cale;{\cal M}^{\rm cond}) } ~~~~~~ \nonumber \\ & \leq &
\sup_{\cale',{\cal N}^{\rm cond}}
 \, I^{q}_{\Phi_1 \otimes \Phi_2 \otimes \cdots \otimes \Phi_{n-1}}
(\cale';{\cal N}^{\rm cond}) +  C_{\shan}(\Phi_n) .
\ee
Iterating (\ref{ineq:four}) gives
\be\label{ineq:five}
\sup_{\cale,{\cal M}^{\rm cond}}
 \, I^{q}_{\Phi_1 \otimes \Phi_2 \otimes \cdots \otimes \Phi_n}
(\cale;{\cal M}^{\rm cond}) \leq
\sum_{k=1}^{n} C_{\shan}(\Phi_k) .
\ee
The definition of conditional capacity is
\be\label{def:Iqcond}
C_{EP}^{\rm cond}(\Phi) =
\lim_{n \rightarrow \infty} \,\frac{1}{n} \,
 \sup_{\cale,{\cal M}^{\rm cond}}
 \, I^{q}_{{\Phi}^{\otimes n}}
(\cale,{\cal M}^{\rm cond}) .
\ee

Hence if we let $\Phi_k = \Phi, ~(k=1,2 \ldots)$
it follows immediately from (\ref{ineq:five}) that
\be \label{eq:final}
C_{EP}^{\rm cond}(\Phi) \leq C_{\shan}(\Phi) .
\ee
Since the capacity of the product channel is never less than the sum of
the channel capacities, i.e,
$C_{EP}^{\rm cond}(\Phi) \geq C_{\shan}(\Phi)$
we must have equality in (\ref{eq:final}) which proves Theorem
\ref{thm:cond}.

It is worth noting that our argument can be used to prove
a somewhat stronger result, namely the additivity of
$\sup_{\cale} I^{q}_{\Phi} (\cale;\calm^{\rm cond}\big)$
for any fixed conditional measurement $\calm^{\rm cond}$.

\medskip

All that remains is to verify  (\ref{eq:mutinfo.ident}) which
is, except for notation, equivalent to
the following result from classical information theory: for any
random variables $J,B,C$
\be\label{eq:three channels}
I^c(J;B,C) = I^c(J;B) + I^c(J;C|\,B\,) .
\ee
Although the derivation of (\ref{eq:three channels}) is quite
elementary (see for example \cite{Cov,NC}), for completeness we include
it in Appendix A, where we also show its equivalence to
(\ref{eq:mutinfo.ident}).

\bigskip

\noindent {\bf Acknowledgment:}  It is a pleasure to thank C.H. Bennett,
J.A. Smolin and B.M. Terhal for useful discussions which helped to
crystallize our understanding of this problem, and
P. Shor for communicating his independent proof of Theorem \ref{thm:cond}.
We are also grateful to the referee for an extremely careful reading
of the previous version.

\appendix

\section{Appendix:  A Useful Information Identity.}

First we relate (\ref{eq:mutinfo.ident}) to an expression involving
classical mutual information. The input alphabet of the product channel
can be described by a classical discrete random variable $J$, whose
distribution is given by the input ensemble ${\cal E}_{12}$,
that is $P(J\!\!=\!\!j) = {\pi}_j$. The output alphabet can be described
similarly  by a
pair of random variables $B,C$, corresponding to the joint POVM
$\wh{\calm}_{12}$. The joint distribution of $J,B,C$ is given by
application of the formula (\ref{eq:dual}), namely
\be\label{eq:app1}
P(J\!\!=\!\!j,B\!\!=\!\!b,C\!\!=\!\!c) = p(j,b,c) = \pi_j \tr [\, (\rho_j)\,\,
F_{b} \otimes F_{c}^{(b)} \,] .
\ee
Applying the definitions in (\ref{eq:IqEM}), (\ref{eq:mut.info.c})
and (\ref{eq:dual}) gives directly
\be
I^c(J;B,C) = I^q(\cale_{12};\wh{\calm}_{12}) .
\ee
Furthermore, by summing over $c$ in (\ref{eq:app1}) and conditioning
on $j$, it follows that
\be
p(b|j) = \tr [\, (\rho_j)\,\, F_{b} \otimes I \,]
= \tr [\, (\rho_j)^{(1)} \,\, F_{b} \,] .
\ee
Comparing with the definition of the ensemble ${\cal E}_{1}$, it
follows that
\be
I^c(J;B) = I^q(\cale_1;\wh{\calm}_1) .
\ee
For the second term on the right side of (\ref{eq:three channels}), recall
that by definition
\be
I^q(J;C\,|\,B) = \sum_{b} p(b) I^q(J;C\,|\,\{B\!=\!b\}) .
\ee
Also
\be
p(c|j,b) = {p(j,b,c) \over p(j,b)} =
\tr [\, (\rho_{j,b})^{(2)} \,\, F^{(b)}_{c} \,]
\ee
and $p(j|b) = p(j,b) / p(b) = p(b|j) {\pi}_j / p(b)$, so therefore
\be
I^q(J;C\,|\,\{B\!=\!b\}) = I^q({\cal E}_{2}(b);\wh{\calm}_{2}(b)) .
\ee
Hence equations
(\ref{eq:mutinfo.ident})
and (\ref{eq:three channels}) are identical.

As noted before, (\ref{eq:three channels}) is a standard result in information
theory. We include its derivation for completeness. The left side can be
rewritten as
\be\label{eq:app2}
I(J;B,C) = H(J) + H(B,C) - H(J,B,C)
\ee
where $H(X)$ is the classical entropy of the random variable $X$.
The two terms on the right side are respectively
\be\label{eq:app3}
I(J;B) = H(J) + H(B) - H(J,B)
\ee
\be\label{eq:app4}
I(J;C\,|\,B) = H(J|B) + H(C|B) - H(J,C|B) .
\ee
Further, for any random variables $X$ and $Y$,
\be\label{eq:app5}
H(X|Y) = H(X,Y) - H(Y),
\ee
and therefore (\ref{eq:app4}) can be written as
\be\label{eq:app6}
I(J;C\,|\,B) = H(J,B) - H(B) + H(C,B) - H(B) - H(J,C,B) + H(B) .
\ee
Adding (\ref{eq:app3}) and (\ref{eq:app6}) gives the right side
of (\ref{eq:app2}), which proves the result.

\bigskip


{~~}

\end{document}